# Chemical versus physical pressure effects on the structure transition of bilayer nickelates

Gang Wang[1,2#], Ningning Wang[1,2#*], Tenglong Lu[1,2#], Stuart Calder[3], Jiaqiang Yan[4], Lifen Shi[1], Jun Hou[1,2], Liang Ma[1,5], Lili Zhang[6], Jianping Sun[1,2], Bosen Wang[1,2], Sheng Meng[1,2,7], Miao Liu[1,7,8*], and Jinguang Cheng[1,2*]

*[1]Beijing National Laboratory for Condensed Matter Physics and Institute of Physics, Chinese Academy of Sciences, Beijing 100190, China*

*[2]School of Physical Sciences, University of Chinese Academy of Sciences, Beijing 100190, China*

*[3]Neutron Scattering Division, Oak Ridge National Laboratory, Oak Ridge, Tennessee 37831, USA*

*[4]Materials Science and Technology Division, Oak Ridge National Laboratory, Oak Ridge, Tennessee 37831, USA*

*[5]Key Laboratory of Materials Physics, Ministry of Education, School of Physics and Microelectronics, Zhengzhou University, Zhengzhou 450052, China*

*[6] Shanghai Synchrotron Radiation Facility, Shanghai Advanced Research Institute, Chinese Academy of Sciences. Shanghai 201204, China*

*[7]Songshan Lake Materials Laboratory, Dongguan 523808, China*

*[8]Center of Materials Science and Optoelectronics Engineering, University of Chinese Academy of Sciences, Beijing 100049, China*

# These authors contribute equally to this work.
*Corresponding authors: nnwang@iphy.ac.cn; mliu@iphy.ac.cn; jgcheng@iphy.ac.cn

The observation of high-$T_c$ superconductivity (HTSC) in concomitant with pressure-induced orthorhombic-tetragonal structural transition in the bilayer La$_3$Ni$_2$O$_7$ has sparked hopes of achieving HTSC by stabilizing the tetragonal phase at ambient pressure. To mimic the effect of external physical pressures, the application of chemical pressure via replacing La$^{3+}$ with smaller rare-earth $R^{3+}$ has been considered as a potential route. Here we clarify the distinct effects of chemical and physical pressures on the structural transition of bilayer nickelates through a combined experimental and theoretical investigation. Contrary to general expectations, we find that substitutions of smaller $R^{3+}$ for La$^{3+}$ in La$_{3-x}R_x$Ni$_2$O$_{7-\delta}$, despite of an overall lattice contraction, produce stronger orthorhombic structural distortions and thus require higher pressures to induce the structural transition. We established a quantitative relationship between the critical pressure $P_c$ for structural transition and the average size of $A$-site cations, $<r_A> \equiv [(3-x) r_{La} + x r_R]/3$. A linear extrapolation of $P_c$ versus $<r_A>$ yields a putative critical value of $<r_A>_c \approx 1.23$ Å for $P_c \approx 1$ bar. The negative correlation between $P_c$ and $<r_A>$ indicates that it is unlikely to reduce $P_c$ to ambient by replacing La$^{3+}$ with smaller $R^{3+}$ ions. Instead, partial substitution of La$^{3+}$ with larger cations such as alkaline-earth Sr$^{2+}$ or Ba$^{2+}$ might be a feasible approach. Our results provide valuable guidelines in the quest of ambient-pressure HTSC in bilayer nickelates.

## Introduction

The recent discovery of high-$T_c$ superconductivity (HTSC) in pressurized La$_3$Ni$_2$O$_7$ stands out as a conspicuous breakthrough in the realm of $3d$ transition-metal oxides, and thus has immediately emerged as a central subject in the community of condensed matter physics [1-28]. As a member of Ruddlesden-Popper (R-P) phase materials, La$_3$Ni$_2$O$_7$ adopts an orthorhombic *Amam* structure at ambient pressure, which consists of alternating layers of La-O and Ni-O planes with a bilayer corner-sharing Ni$^{2.5+}$O$_6$ octahedra structure [29-31]. Recent studies showed that it undergoes an orthorhombic to tetragonal structural transition at high pressure above $P_c$ ~14 GPa, which was believed to play a decisive role in the generation of HTSC [1]. When the higher symmetry *I4/mmm* structure is induced by pressure, the out-of-plane Ni-O-Ni bond angle within the bilayer corner-sharing Ni$^{2.5+}$O$_6$ octahedra is changed to 180° [1, 6, 24]. This will cause the $3d_{z^2}$ orbital to intersect with Fermi level, resulting in the emergency of the $\gamma$ band and promoting superconducting pairing [1]. Such a physical scenario was supported by subsequent theoretical calculations [9, 11, 12, 20, 32, 33], and was employed to explain the bulk superconductivity in polycrystalline La$_2$PrNi$_2$O$_7$ samples with significantly reduced intergrowth of various R-P phases [5].

As an emerging novel class of pressure-induced high-$T_c$ superconductors, it is highly desirable to explore the possibility of achieving superconductivity at ambient pressure. One possible approach is to replace La$^{3+}$ with either isovalent smaller-size rare-earth $R^{3+}$ ions or heterovalent cations. While the latter affects both the lattice and charge balance, the former solely introduces chemical pressure, which is expected to contract the lattice and reduce the physical pressure required to induce structural transition and HTSC. We are thus motivated to investigate the structural evolution of a series of rare-earth substituted La$_{3-x}$$R_x$Ni$_2$O$_{7-\delta}$ samples under high pressure, aiming to reveal the relationship between the critical pressure $P_c$ and the average ionic radius of the $A$-site cations, $<r_A> \equiv [(3-x)\, r_{La} + x\, r_R]/3$. Considering the significant challenges in preparing single crystals of this system, we focus on polycrystalline samples that can be prepared in a controlled manner and are conducive to sample screening before committing to dedicated crystal growth endeavor [2, 5, 29].

In this work, we prepared a series of La$_{3-x}$$R_x$Ni$_2$O$_{7-\delta}$ ($R$ = Pr, Nd, Tb) polycrystalline samples with the same procedures and conducted systematic high-pressure synchrotron XRD (HP-SXRD) to determine the critical pressure $P_c$ for the structural transition. The constructed phase diagram of $P_c$ versus $<r_A>$ reveals that $P_c$ rises monotonically with the reduction of $<r_A>$, which is attributed to the enhanced orthorhombic distortion upon reducing $<r_A>$. This observation indicates that it is unlikely to reduce $P_c$ through the substitution of smaller rare-earth elements at the La position. In addition, the extrapolation of $P_c$ versus $<r_A>$ suggests that it is likely to reduce $P_c$ to ambient pressure via substituting $La^{3+}$ with larger heterovalent cations, such as alkaline-earth $Sr^{2+}$ or $Ba^{2+}$. By unveiling a negative relationship between $<r_A>$ and $P_c$, our results provide valuable insights in achieving superconductivity at ambient pressure in this system.

## Methods

### Sample synthesis

A series of $La_{3-x}R_xNi_2O_{7-\delta}$ ($R$ = Pr, Nd, Tb) polycrystalline samples were synthesized by the sol-gel method as described in Ref [2, 29]. Stoichiometric mixtures of rare-earth oxides and $Ni(NO_3)_2 \cdot 6H_2O$ (99.99%, Alfa Aesar) were used as the starting materials. All samples were synthesized using the identical procedures and sintered at the same temperature conditions. Each dopant has its own solubility, which we determined by experimenting with different compositions. In this work, the highest experimentally obtained doping level of $La_{3-x}R_xNi_2O_{7-\delta}$ ($R$ = Pr, Nd, Tb) samples with a pure phase are $La_2PrNi_2O_{7-\delta}$, $La_{1.8}Nd_{1.2}Ni_2O_{7-\delta}$, and $La_{2.7}Tb_{0.3}Ni_2O_{7-\delta}$. Further increasing the substitution content results in the formation of $(La, R)_2NiO_4$ or other oxide impurities.

### Structural characterizations

The phase purity and crystal structure of $La_{3-x}R_xNi_2O_{7-\delta}$ at ambient conditions were determined by powder X-ray diffraction (XRD) collected via a Huber diffractometer with Cu-$K_\alpha$ radiation. Neutron powder diffraction (NPD) measurements were carried out using the HB-2A diffractometer at the High Flux Isotope Reactor (HFIR) of Oak Ridge National Laboratory (ORNL) [34, 35]. Powder samples of $La_{3-x}R_xNi_2O_{7-\delta}$ were placed inside a 6 mm-diameter vanadium container and then inserted into a closed cycle refrigerator. NPD data was collected at 295 K, utilizing a constant wavelength of $\lambda$ = 1.5365 Å, derived from the Ge (115) monochromator. The NPD pattern was collected by scanning a 120° bank of 44 $^3$He detectors in 0.05° steps to give $2\theta$ coverage from 5° to 150°. Rietveld refinements were performed with the FULLPROF program [36]. HP-SXRD measurements were performed at BL15U1 station of Shanghai Synchrotron Radiation Facility (SSRF) with a wavelength of $\lambda$ = 0.6199 Å. Rietveld analysis was performed with GSAS-II program [37].

### DFT structure calculations

Density functional theory (DFT) calculations were conducted using the projector-augmented wave (PAW) method, implemented in the Vienna *ab initio* simulation package (VASP) [38, 39]. All calculations utilized the Perdew-Burke-Ernzerhof (PBE) generalized gradient approximation (GGA) exchange-correlation functional [40]. To correct the self-interaction error on the Ni species, a rotationally averaged Hubbard $U$ correction of 3 eV was applied [41]. This value is commonly used in theoretical studies of this system [7, 9, 15, 19, 42] and is close to the value of ~3.5 eV obtained from the angle-resolved photoelectron spectroscopy experiments. [21] For all calculations, a plane wave energy cutoff of 520 eV, an electronic minimization threshold of $10^{-6}$ eV, and a $k$-point grid of $n_{kpoints} \times n_{atoms} > 1000$ were adopted. For structural relaxation, all degrees of freedom of atoms and lattices were allowed to relax until the atomic forces were less than 0.001 eV Å$^{-1}$. Given that La$_3$Ni$_2$O$_7$ displays no long-range magnetic ordering in previous experiments [1], all DFT calculations in this study were performed without spin-polarization. The PSTRESS parameter was employed to control the external hydrostatic pressure.

## Results and discussion

The phase purity of obtained La$_{3-x}$$R_x$Ni$_2$O$_{7-\delta}$ samples was first examined with XRD at room temperature. As shown in Fig. S1, all samples were confirmed to be singe phase with the orthorhombic *Amam* structure (No. 63). The substitution of smaller $R^{3+}$ for La$^{3+}$ in the perovskite-type La$_3$Ni$_2$O$_7$ is expected to shrink the lattice and enhance local structural distortions. To obtain reliable information about the structural changes upon different $R^{3+}$ substitutions, we performed NPD measurements at ambient conditions since the oxygen atoms have a large neutron scattering length. Fig. 1(a) and Table S1 display the Rietveld refinement results on the collected NPD data, which further confirm that all samples are single phase with the orthorhombic structure. The obtained lattice parameters are displayed in Fig. 1(b, c) as a function of the average size of $A$-site cations, <$r_A$>. As can be seen, the lattice parameters exhibit anisotropic shrinkage with increasing <$r_A$>, i.e., $a$ and $c$ are reduced monotonically by 0.6% and 0.9%, respectively, while $b$ remains nearly unchanged, resulting in a net volume reduction of 1.5% from La$_3$Ni$_2$O$_{7-\delta}$ to La$_{1.8}$Nd$_{1.2}$Ni$_2$O$_{7-\delta}$. Such an evolution confirms that we have successfully introduced some chemical pressure into the La$_3$Ni$_2$O$_7$ lattice by replacing La$^{3+}$ with smaller $R^{3+}$.

Although the chemical and physical pressures play similar roles in shrinking the overall lattice, their influences on the local structural distortions are distinct. For example, the external physical pressure usually compresses the lattice uniformly and drives it to a higher symmetry, while the chemical pressure produces stronger local structural distortions in the perovskite-type structures. As shown in Fig. 1(d, e), some interesting features about the local structural modifications induced by smaller $R^{3+}$ are noteworthy: (1) both the out-of-plane Ni-O2/O1 bonds first experience an elongation, reaching the



maximum at $<r_A> = 1.206$ Å, and then decreases upon further reducing $<r_A>$ for La$_{1.8}$Nd$_{1.2}$Ni$_2$O$_7$; (2) the in-plane Ni-O3 and Ni-O4 bond lengths vary oppositely, resulting in reduction first and then an enhanced splitting of in-plane Ni-O bonds as a function of $<r_A>$; (3) both in-plane and out-of-plane buckling of the corner-shared NiO$_6$ octahedra becomes stronger as the Ni-O-Ni bond angles are reduced. Therefore, the substitutions of smaller $R^{3+}$ for La$^{3+}$ in La$_{3-x}$R$_x$Ni$_2$O$_{7-\delta}$ not only shrink the overall lattice but also produce stronger local orthorhombic distortions. The former effect is similar as external physical pressure, whereas the latter one should require higher pressure to induce a structural transition to a higher symmetry.

To determine the critical pressure $P_c$ for the structural transition, we conducted HP-SXRD measurements at room temperature under various pressures up to 40 GPa. The HP-SXRD results on two representative samples, La$_3$Ni$_2$O$_{7-\delta}$ and La$_{1.8}$Nd$_{1.2}$Ni$_2$O$_{7-\delta}$, are depicted in Fig. 2(a) and (b), respectively. For La$_3$Ni$_2$O$_{7-\delta}$, the SXRD patterns below 4.9 GPa consistently match the orthorhombic *Amam* structure, as shown by the representative refinement at 3.6 GPa, Fig. 2(c). However, upon compression to 7.0 GPa, several adjacent peaks merge, such as the (020) and (200) peaks at ~13.4°, as illustrated in the right panel of Fig. 2(a). This observation suggests the occurrence of pressure-induced structural transition towards a higher symmetry. Subsequent structure analyses revealed that the SXRD patterns of the HP phase can be better described using the Sr$_3$Ti$_2$O$_7$-type structural model with the tetragonal *I4/mmm* space group (No. 139), Fig. 2(d), consistent with previously reported results [24]. For La$_{1.8}$Nd$_{1.2}$Ni$_2$O$_{7-\delta}$, the same pressure-induced orthorhombic to tetragonal structural transition was also observed, with the critical pressure $P_c \approx 17$ GPa, Fig. 2(b, e, f). The HP-SXRD results for the other three samples of the La$_{3-x}$R$_x$Ni$_2$O$_{7-\delta}$ ($R$ = Pr, Nd, Tb) series were shown in Figs. S2-S4. They all undergo the same structural phase transition under high pressure with the $P_c \approx 8$, 10.5, and 11 GPa for La$_{2.7}$Pr$_{0.3}$Ni$_2$O$_{7-\delta}$, La$_{2.7}$Tb$_{0.3}$Ni$_2$O$_{7-\delta}$, and La$_2$PrNi$_2$O$_{7-\delta}$, respectively.

Fig. 3 displays the lattice parameters as a function of pressure for these two samples extracted from their HP-SXRD patterns after Rietveld refinements. As seen from Fig. 3(a,b), their lattice parameters decrease continuously with increasing pressure, but exhibit anisotropic compressions. In the lower pressure range, lattice parameter $b$ decreases faster than $a$ and they converge at $P_c$, where the structural transition takes place. As the crystal structure transforms into a higher symmetry tetragonal structure, the lattice parameter $a$ contracts by a factor of $\frac{1}{\sqrt{2}}$, leading to a 0.5 times reduction in the unit-cell volume, $V$. The collected pressure-volume $P(V)$ data in the whole pressure range can be fitted to the third-order Birch-Murnaghan equation[43], yielding a bulk modulus of $B_0^{orth} = 137.1$ GPa and $B_0^{tetra} = 208.4$ GPa for La$_3$Ni$_2$O$_{7-\delta}$ and $B_0^{orth} = 179.3$



GPa and $B_0^{\text{tetra}}$ = 221.7 GPa for La$_{1.8}$Nd$_{1.2}$Ni$_2$O$_{7-\delta}$, respectively, with B$_0^{'}$ fixed at 3.5, as shown by the dashed lines in Fig. 3(c, d).

From the above experimental results, we can see that $P_c$ increases with reducing $<r_A>$. To understand the structural transition of bilayer nickelates as a function of chemical and physical pressures, we employed the DFT calculations to study the structural evolution of the $R_3$Ni$_2$O$_{7-\delta}$ ($R$ = La, Pr, Nd) samples under high pressure. Fig. 4 shows the calculated results of lattice parameters for $R_3$Ni$_2$O$_{7-\delta}$ ($R$ = La, Pr, Nd) as a function of the external pressure. As can be seen, all $R_3$Ni$_2$O$_7$ compounds undergo an orthorhombic to tetragonal structural transition under high pressure, consistent with the experimental observations mentioned above. Additionally, our DFT calculations show that the calculated $P_c$ at $U$ = 3eV for La$_3$Ni$_2$O$_7$ is ~9 GPa, which is close to the HP-SXRD observations, indicating the reliability and accuracy of our theoretical methods. It should be noted that the critical pressure $P_c$ from DFT calculations depends sensitively on the choice of Hubbard $U$ values, and a higher $U$ value results in a lower calculated $P_c$. As shown in Fig. S5, the calculated $P_c$ of La$_3$Ni$_2$O$_7$ decreases monotonically from ~ 12 GPa for $U$ = 1 eV to ~ 5 GPa for $U$ = 6 eV. The $U$ value of 3 eV applied in this work is favored by many theoretical studies of this system [7, 9, 15, 19, 42] and is close to the value of ~3.5 eV obtained from the angle-resolved photoelectron spectroscopy experiments. [21]

To better clarify the impact of physical pressure on the structural evolution of the system, we extracted the bond lengths and bond angles of La$_3$Ni$_2$O$_7$ under various pressures from DFT calculations. As shown in Fig. S6(a), all bond lengths decrease progressively with applied pressure but exhibit anisotropic compressions. For the two out-of-plane bonds, Ni-O2 shortens more rapidly than Ni-O1 with increasing pressure, leading to a reduction in the length difference between these bonds. For the in-plane bonds, Ni-O4 decreases in length faster than Ni-O3, and they converge at the critical pressure $P_c$. In addition, as the Ni–O–Ni bond angle gradually increases with pressure and reaches 180° at $P_c$, both in-plane and out-of-plane buckling of the corner-shared NiO$_6$ octahedra are diminished, Fig. S6(b). These results suggest that the effects of physical pressure on structural evolution differ significantly from those induced by chemical pressure, as discussed above.

Then, we plotted in Fig. 5 the obtained $P_c$ of La$_{3-x}R_x$Ni$_2$O$_{7-\delta}$ from both experiments and DFT calculations as a function of $<r_A>$. As seen clearly, $P_c$ increases with the reduction of $<r_A>$. This result reveals that the structural transition is dictated mainly by the local orthorhombic distortions in chemically pre-compressed La$_{3-x}R_x$Ni$_2$O$_{7-\delta}$ samples; i.e., the stronger local structural distortion, the higher $P_c$ for structural transition. The observed evolution of $P_c$ versus $<r_A>$ implies that it is not feasible to reduce $P_c$ to



ambient pressure via substituting $La^{3+}$ with smaller-size $R^{3+}$ ions. A linear extrapolation of $P_c$ vs $<r_A>$ to zero would yield an $<r_A>_c \approx 1.232$ Å for the stabilization of the tetragonal phase at ambient pressure, as shown by the black dashed line in Fig. 5. If this assumption holds, alkaline-earth-metal substituted $La_{2.5}Sr_{0.5}Ni_2O_7$ and $La_{2.8}Ba_{0.2}Ni_2O_7$ can satisfy such a requirement of $<r_A> \approx 1.232$ Å. Although the replacement of $La^{3+}$ with heterovalent alkaline-earth cations could be a viable approach to stabilize the tetragonal phase at ambient pressure, it remains quite challenging to synthesize these samples.

**Discussions**

Recently, Jiao et al. [18] synthesized a series of Sr-doped $La_{3-x}Sr_xNi_2O_{7-\delta}$ ($0 \leq x \leq 0.1$) polycrystalline samples via conventional solid-state-reaction method. They found that upon increasing x, the lattice constants $a$ and $b$ expand while the $c$-axis contracts, with a simultaneous enhancement of Ni-O-Ni bond angles. In addition, the resistivity decreases monotonically in the entire temperature range due to the introduction of hole carriers and the reduced structural distortions. Xu et al. [44] successfully obtained Sr-doped $La_{2.8}Sr_{0.2}Ni_2O_{6.95}$ single crystals with the orthorhombic bilayer structure by treating the precursors at 20 GPa and 1400 °C in a Walker-type multianvil press. For such samples, both out-of-plane Ni–O–Ni bond angles of 173.4(2)° and in-plane Ni-O-Ni bond angles of 175.0(2)° and 176.7(2)° are indeed larger than those of pristine $La_3Ni_2O_7$. The HP-SXRD was not performed on those $La_{3-x}Sr_xNi_2O_{7-\delta}$ polycrystalline and $La_{2.8}Sr_{0.2}Ni_2O_{6.95}$ single crystal. According to the present study, it is expected that the structural transition to the tetragonal phase should take place at a lower $P_c$ than $La_3Ni_2O_7$. Thus, a HP structural study is highly desirable. Intriguingly, it was found that the $La_{2.8}Sr_{0.2}Ni_2O_{6.95}$ single crystal exhibits an insulating behavior at ambient pressure and displays pressure-driven insulator-metal-insulator crossovers up to 19 GPa. These transport properties are dramatically different from those of the parent compound $La_3Ni_2O_{7-\delta}$ and warrant further investigation. Nevertheless, the successful synthesis of Sr-doped $La_{2.8}Sr_{0.2}Ni_2O_{6.95}$ single crystals would open an avenue for exploring possible ambient-pressure HTSC via further enhancing the substitution levels of $Sr^{2+}$ for $La^{3+}$ in the bilayer nickelates.

It is noteworthy that the critical pressure $P_c \approx 7$ GPa for our polycrystalline $La_3Ni_2O_{7-\delta}$ sample is lower than that of ~14 GPa for the single-crystal samples reported by Sun *et al* in reference [1]. At present, the origin of this discrepancy remains unclear, which might arise from the distinct grain sizes or diverse levels of structure defects/disorders for the different forms of samples. Nonetheless, our HP-SXRD results on the polycrystalline $La_{3-x}R_xNi_2O_{7-\delta}$ samples prepared at the same conditions should provide self-consistent information about the effect of substitution of smaller $R^{3+}$ ions at the La sites as discussed above.



## Conclusion

In summary, we conducted systematic HP-SXRD measurements on a series of $La_{3-x}R_xNi_2O_{7-\delta}$ ($R$ = Pr, Nd, Tb) samples and established a quantitative relationship between the critical pressure $P_c$ for orthorhombic-tetragonal structural transition and the average size of $A$-site cations, $<r_A>$. By unveiling the inverse relationship between $P_c$ and $<r_A>$, our results provide useful guidelines to achieve the tetragonal phase at ambient pressure in this system. Replacing La with alkaline-earth metals can increase $<r_A>$ and introduce additional charge carriers, making it a promising approach to achieve HTSC at ambient pressure.

## Acknowledgments


This work is supported by the National Key Research and Development Program of China (2023YFA1406100, 2021YFA1400200), the National Natural Science Foundation of China (12025408, 11921004, U23A6003), the Strategic Priority Research Program of CAS (XDB33000000), the Postdoctoral Fellowship Program of China Postdoctoral Science Foundation (GZB20230828), the China Postdoctoral Science Foundation (2023M743740), CAS PIFI program (2024PG0003), and CAS Project for Young Scientists in Basic Research (2022YSBR-047). JQY was supported by the U.S. Department of Energy, Office of Science, Basic Energy Sciences, Division of Materials Sciences and Engineering. High-pressure synchrotron X-ray measurements were performed at the BL15U1 station of the Shanghai Synchrotron Radiation Facility (SSRF), which is supported by the Chinese Academy of Sciences. This research used resources at the High Flux Isotope Reactor, a U.S. DOE Office of Science User Facility operated by the Oak Ridge National Laboratory.


## Author contributions

J.G.C. designed the project. G.W. and N.N.W. synthesized the materials and characterized their structure via XRD; G.W., N.N.W., L.F.S., J.H. and L.M. measured the HP-SXRD and analyzed the data with the support of J.P.S., B.S.W. and L.L.Z.; C. S. and J.Q.Y. measured and analyzed the NPD data; T.L.L. performed the DFT structure calculations with the support of S.M., and M.L.; J.G.C., N.N.W. and G.W. wrote the paper with inputs from all coauthors.

## Competing interests

The authors declare no competing interests.

## Reference




1.  Sun H, *et al.* Signatures of superconductivity near 80 K in a nickelate under high pressure. *Nature* **621**, 493-498 (2023).

2.  Wang G, *et al.* Pressure-induced superconductivity in polycrystalline $La_3Ni_2O_{7-\delta}$. *Phys Rev X* **14**, 011040 (2024).

3.  Zhang Y, *et al.* High-temperature superconductivity with zero resistance and strange-metal behaviour in $La_3Ni_2O_{7-\delta}$. *Nat Phys*, in press (2024).

4.  Hou J, *et al.* Emergence of high-temperature superconducting phase in the pressurized $La_3Ni_2O_7$ crystals. *Chin Phys Lett* **40**, 117302 (2023).

5.  Wang N, *et al.* Bulk high-temperature superconductivity in the highpressure tetragonal phase of bilayer $La_2PrNi_2O_7$. *arXiv:240705681*, (2024).

6.  Wang M, Wen H-H, Wu T, Yao D-X, Xiang T. Normal and superconducting properties of $La_3Ni_2O_7$. *Chin Phys Lett* **41**, 077402 (2024).

7.  Sakakibara H, Kitamine N, Ochi M, Kuroki K. Possible high $T_c$ superconductivity in $La_3Ni_2O_7$ under high pressure through manifestation of a nearly half-filled bilayer hubbard model. *Phys Rev Lett* **132**, 106002 (2024).

8.  Qu X-Z, *et al.* Bilayer $t-J-J_\perp$ model and magnetically mediated pairing in the pressurized nickelate $La_3Ni_2O_7$. *Phys Rev Lett* **132**, 036502 (2024).

9.  Luo Z, Hu X, Wang M, Wú W, Yao D-X. Bilayer two-orbital model of $La_3Ni_2O_7$ under pressure. *Phys Rev Lett* **131**, 126001 (2023).

10. Lu C, Pan Z, Yang F, Wu C. Interlayer-coupling-driven high-temperature superconductivity in $La_3Ni_2O_7$ under pressure. *Phys Rev Lett* **132**, 146002 (2024).

11. Liu Y-B, Mei J-W, Ye F, Chen W-Q, Yang F. s$^\pm$-wave pairing and the destructive role of apical-oxygen deficiencies in $La_3Ni_2O_7$ under pressure. *Phys Rev Lett* **131**, 236002 (2023).

12. Jiang R, Hou J, Fan Z, Lang Z-J, Ku W. Pressure driven fractionalization of ionic spins results in cupratelike high-$T_c$ superconductivity in $La_3Ni_2O_7$. *Phys Rev Lett* **132**, 126503 (2024).

13. Christiansson V, Petocchi F, Werner P. Correlated electronic structure of $La_3Ni_2O_7$ under pressure. *Phys Rev Lett* **131**, 206501 (2023).

14. Chen K, *et al.* Evidence of Spin Density Waves in $La_3Ni_2O_{7-\delta}$. *Phys Rev Lett* **132**, 256503 (2024).

15. Rhodes LC, Wahl P. Structural routes to stabilize superconducting $La_3Ni_2O_7$ at ambient pressure. *Phys Rev Mater* **8**, (2024).

16. Li F, *et al.* Design and synthesis of three-dimensional hybrid Ruddlesden-Popper nickelate single crystals. *Phys Rev Mater* **8**, 053401 (2024).



17.    Wú W, Luo Z, Yao D-X, Wang M. Superexchange and charge transfer in the nickelate superconductor La$_3$Ni$_2$O$_7$ under pressure. *Sci China-Phys Mech Astron* **67**, 117402 (2024).

18.    Jiao K, Niu R, Xu H, Zhen W, Wang J, Zhang C. Enhanced conductivity in Sr doped La$_3$Ni$_2$O$_{7-\delta}$ with high-pressure oxygen annealing. *Physica C: Superconductivity and its Applications* **621**, (2024).

19.    Geisler B, Hamlin JJ, Stewart GR, Hennig RG, Hirschfeld PJ. Structural transitions, octahedral rotations, and electronic properties of A$_3$Ni$_2$O$_7$ rare-earth nickelates under high pressure. *npj Quantum Materials* **9**, (2024).

20.    Zhang Y, Lin LF, Moreo A, Maier TA, Dagotto E. Structural phase transition, s$^{\pm}$-wave pairing, and magnetic stripe order in bilayered superconductor La$_3$Ni$_2$O$_7$ under pressure. *Nat Commun* **15**, 2470 (2024).

21.    Yang J, *et al.* Orbital-dependent electron correlation in double-layer nickelate La$_3$Ni$_2$O$_7$. *Nat Commun* **15**, 4373 (2024).

22.    Dong Z, *et al.* Visualization of oxygen vacancies and self-doped ligand holes in La$_3$Ni$_2$O$_{7-\delta}$. *Nature* **630**, 847–852 (2024).

23.    Kakoi M, *et al.* Multiband metallic ground state in multilayered nickelates La$_3$Ni$_2$O$_7$ and La$_4$Ni$_3$O$_{10}$ probed by $^{139}$La-NMR at ambient pressure. *J Phys Soc Jpn* **93**, 053702 (2024).

24.    Wang L, *et al.* Structure eesponsible for the superconducting state in La$_3$Ni$_2$O$_7$ at high-pressure and low-temperature conditions. *J Am Chem Soc* **146**, 7506-7514 (2024).

25.    Chen X, *et al.* Polymorphism in the Ruddlesden–Popper nickelate La$_3$Ni$_2$O$_7$: Discovery of a hidden phase with distinctive layer stacking. *J Am Chem Soc* **146**, 3640-3645 (2024).

26.    Wang H, Chen L, Rutherford A, Zhou H, Xie W. Long-range structural order in a hidden phase of Ruddlesden-Popper bilayer nickelate La$_3$Ni$_2$O$_7$. *Inorg Chem* **63**, 5020-5026 (2024).

27.    Shen Y, Qin M, Zhang G-M. Effective Bi-Layer Model Hamiltonian and Density-Matrix Renormalization Group Study for the High-$T_c$ Superconductivity in La$_3$Ni$_2$O$_7$ under High Pressure. *Chin Phys Lett* **40**, 127401 (2023).

28.    Jiang K, Wang Z, Zhang F-C. High-Temperature Superconductivity in La$_3$Ni$_2$O$_7$. *Chin Phys Lett* **41**, 017402 (2024).

29.    Zhang Z, Greenblatt M, Goodenough JB. Synthesis, structure, and properties of the layered perovskite La$_3$Ni$_2$O$_{7-\delta}$. *J Solid State Chem* **108**, 402-409 (1994).





30. Greenblatt M. Ruddlesden-Popper $Ln_{n+1}Ni_nO_{3n+1}$ nickelates: structure and properties. *CURR OPIN SOLID ST M* **2**, 174-183 (1997).

31. Ling CD, Argyriou DN, Wu G, Neumeier JJ. Neutron diffraction study of $La_3Ni_2O_7$: structural relationships among n=1, 2, and 3 phases $La_{n+1}Ni_nO_{3n+1}$. *J Solid State Chem* **152**, 517-525 (2000).

32. Yang Y-f, Zhang G-M, Zhang F-C. Interlayer valence bonds and two-component theory for high-$T_c$ superconductivity of $La_3Ni_2O_7$ under pressure. *Phys Rev B* **108**, L201108 (2023).

33. Zhang Y, Lin L-F, Moreo A, Maier TA, Dagotto E. Electronic structure, magnetic correlations, and superconducting pairing in the reduced Ruddlesden-Popper bilayer $La_3Ni_2O_6$ under pressure: Different role of $d_{3z^2-r^2}$ orbital compared with $La_3Ni_2O_7$. *Phys Rev B* **109**, 045151 (2024).

34. Garlea VO, *et al.* The high-resolution powder diffractometer at the high flux isotope reactor. *Appl Phys A* **99**, 531-535 (2010).

35. Calder S, *et al.* A suite-level review of the neutron powder diffraction instruments at Oak Ridge National Laboratory. *Rev Sci Instrum* **89**, 092701 (2018).

36. Rodríguez-Carvajal J. Recent advances in magnetic structure determination by neutron powder diffraction. *Physica B: Condensed Matter* **192**, 55-69 (1993).

37. Toby BH, Von Dreele RB. GSAS-II: the genesis of a modern open-source all purpose crystallography software package. *J Appl Crystallogr* **46**, 544-549 (2013).

38. Kresse G, Furthmüller J. Efficiency of ab-initio total energy calculations for metals and semiconductors using a plane-wave basis set. *Comput Mater Sci* **6**, 15-50 (1996).

39. Kresse G, Joubert D. From ultrasoft pseudopotentials to the projector augmented-wave method. *Phys Rev B* **59**, 1758-1775 (1999).

40. Perdew JP, Burke K, Ernzerhof M. Generalized Gradient Approximation Made Simple. *Phys Rev Lett* **77**, 3865-3868 (1996).

41. Dudarev SL, Botton GA, Savrasov SY, Humphreys CJ, Sutton AP. Electron-energy-loss spectra and the structural stability of nickel oxide: An LSDA+U study. *Phys Rev B* **57**, 1505-1509 (1998).

42. Sakakibara H, *et al.* Theoretical analysis on the possibility of superconductivity in the trilayer Ruddlesden-Popper nickelate $La_4Ni_3O_{10}$ under pressure and its experimental examination: Comparison with $La_3Ni_2O_7$. *Phys Rev B* **109**, (2024).

43. Birch F. Finite elastic strain of cubic crystals. *Phys Rev* **71**, 809-824 (1947).





44.     Xu M, *et al.* Pressure-Dependent "Insulator–Metal–Insulator" Behavior in Sr-Doped La$_3$Ni$_2$O$_7$. *Adv Electron Mater*, 2400078 (2024).




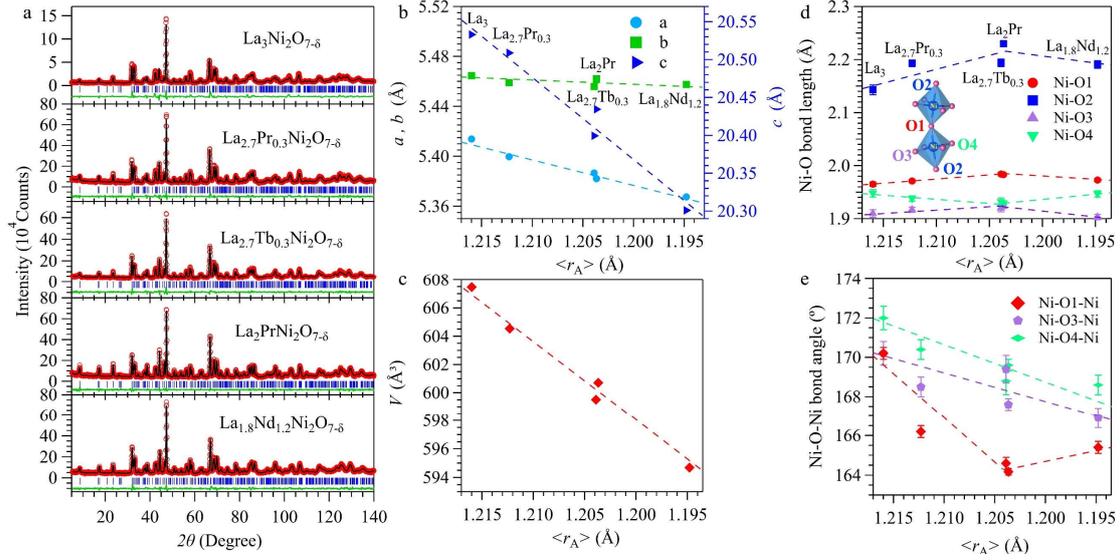

**Figure 1. Structural characterizations of the La$_{3-x}$$R_x$Ni$_2$O$_{7-\delta}$ ($R$ = Pr, Nd, Tb) polycrystalline samples.** (a) Rietveld refinements on the NPD data with the space group *Amam*. (b-e) The obtained lattice parameters, unit-cell volume, Ni-O bond lengths and Ni-O-Ni bond angles as a function of the average ionic radius of the *A*-site cations, <$r_A$>, across the series of La$_{3-x}$$R_x$Ni$_2$O$_{7-\delta}$ ($R$ = Pr, Nd, Tb) polycrystalline samples. The NPD data for La$_{3-x}$Pr$_x$Ni$_2$O$_{7-\delta}$ (x = 0, 1.0) samples were adopted from our previous study [2, 5].



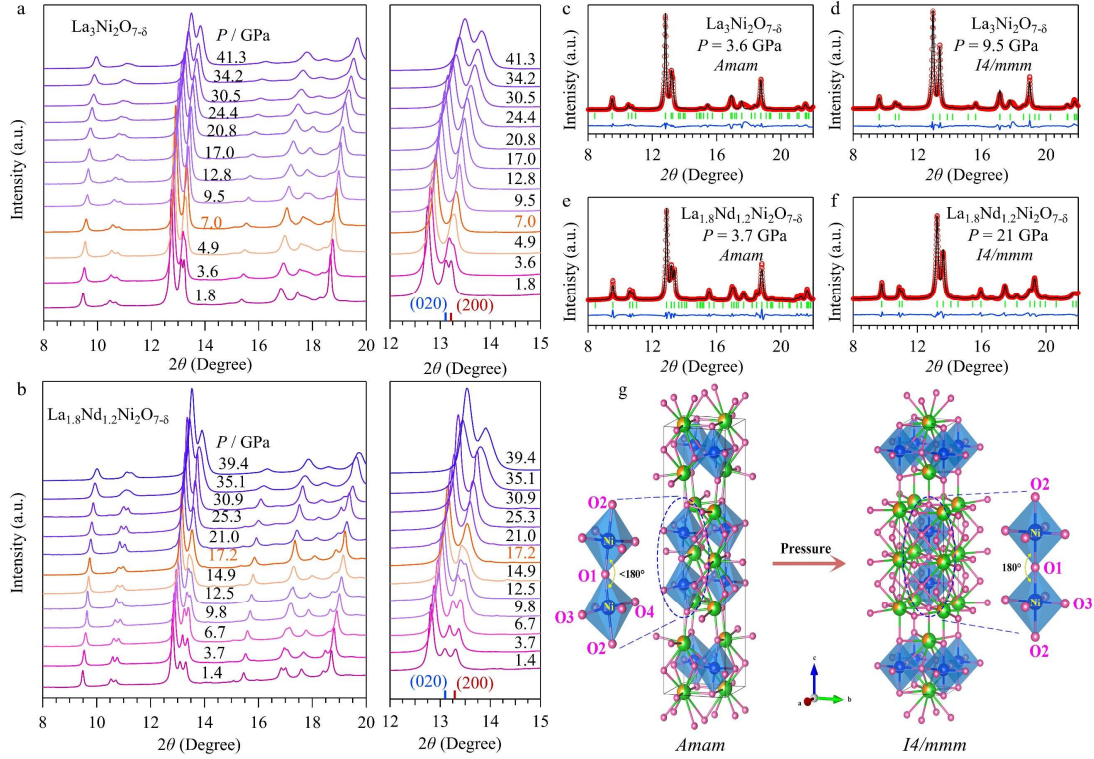

**Figure 2. Pressure-induced structural transition.** (a, b) SXRD patterns of La$_3$Ni$_2$O$_{7-\delta}$ and La$_{1.8}$Nd$_{1.2}$Ni$_2$O$_{7-\delta}$ polycrystalline samples under various pressures up to 41.3 and 39.4 GPa, respectively. The enlarged view of SXRD around the representative $2\theta$ ranges highlight the gradual merging of the diffraction peaks upon compression. (c-f) Refinement results of the SXRD patterns at pressures before and after the structural phase transition for La$_3$Ni$_2$O$_{7-\delta}$ and La$_{1.8}$Nd$_{1.2}$Ni$_2$O$_{7-\delta}$. (g) Crystal structure transformation of La$_{3-x}$$R_x$Ni$_2$O$_{7-\delta}$ under high pressure.

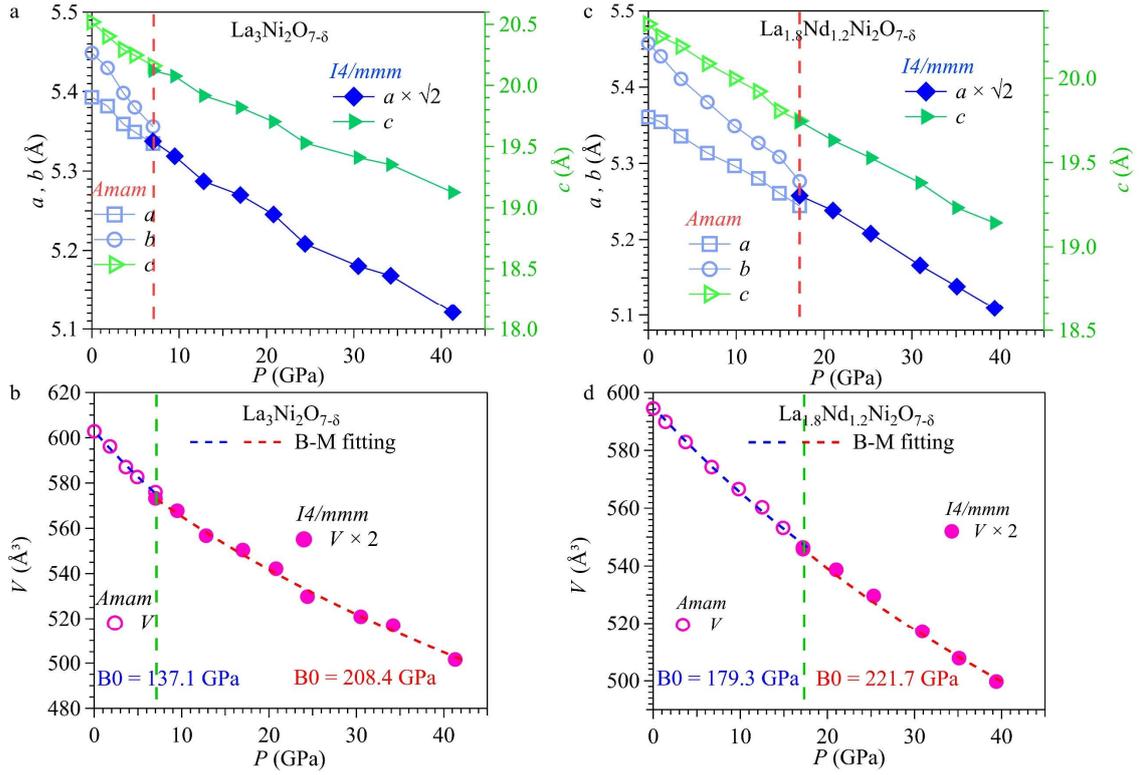

**Figure 3. The evolution of lattice parameters under high pressure for La₃Ni₂O₇₋δ and La₁.₈Nd₁.₂Ni₂O₇₋δ polycrystalline samples.** Lattice parameters and cell volume as a function of pressure for (a, b) La₃Ni₂O₇₋δ and (c, d) La₁.₈Nd₁.₂Ni₂O₇₋δ samples under various pressures up to 41.3 and 39.4 GPa, respectively. The critical pressure $P_c$ for the structural transition is marked by the vertical broken line in (b, d).

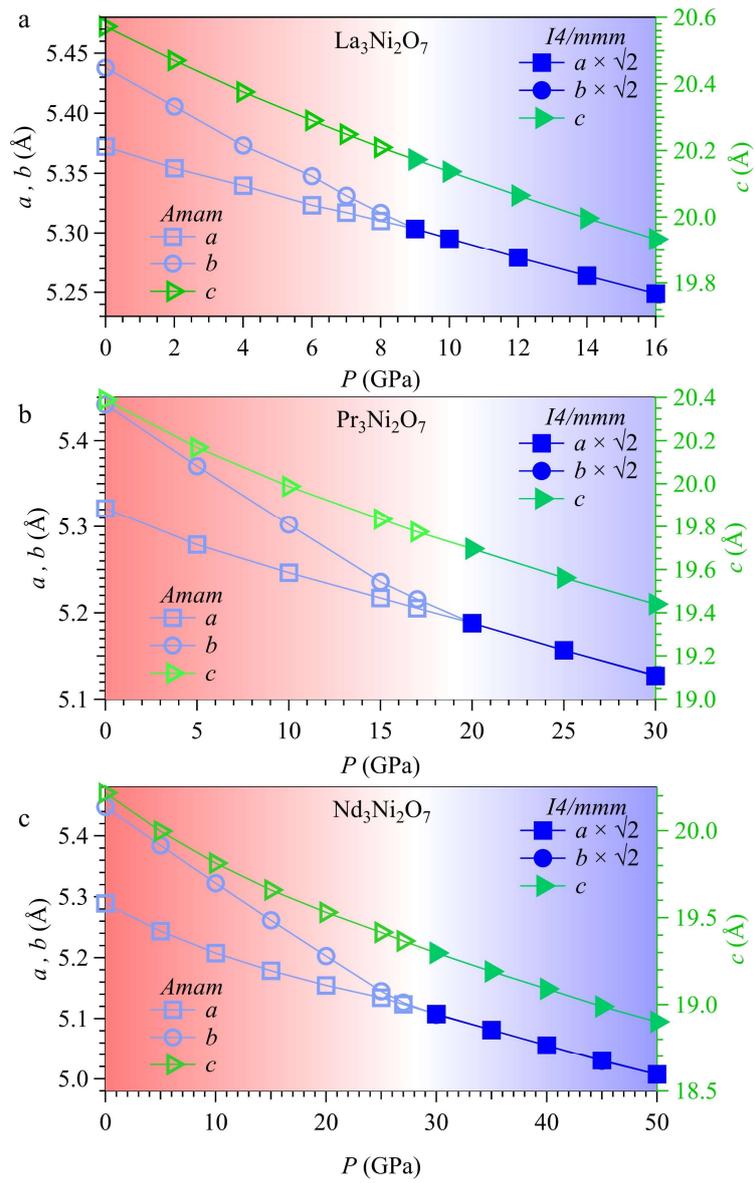

**Figure 4. The calculated structural evolutions as a function of pressure.** (a-c) Lattice parameters of $R_3Ni_2O_7$ ($R$ = La, Pr, Nd) as a function of the external pressure from DFT calculations. As can be seen, the lattice parameters $a$ and $b$ of all $R_3Ni_2O_7$ compounds gradually converge under high pressure, but with different critical pressures.

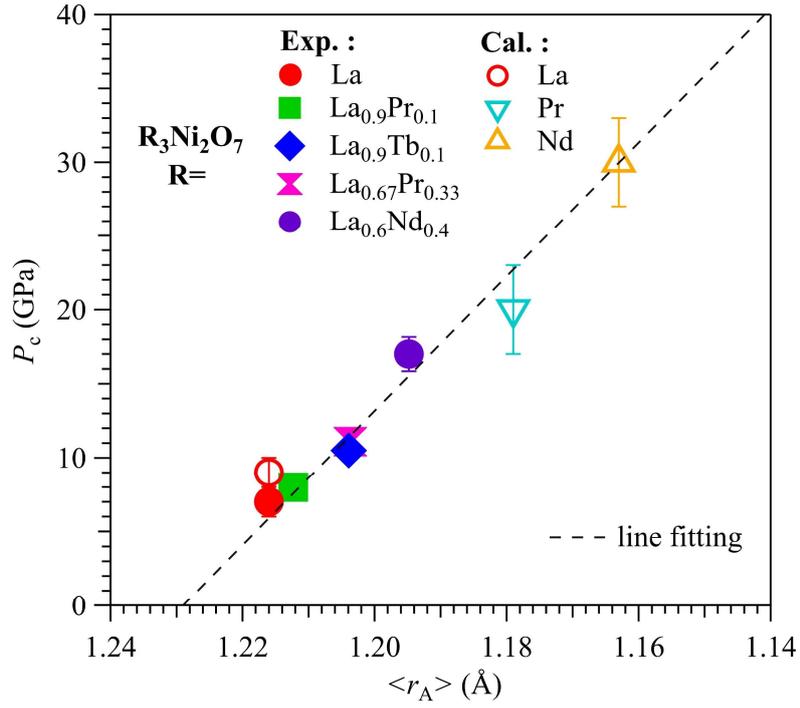

**Figure 5. The constructed $P_c$ -$<r_A>$ phase diagram.** The critical pressure $P_c$ for the orthorhombic-tetragonal structural transition as a function of the average ionic radius of the $A$-site cations, $<r_A>$, across the series of $La_{3-x}R_xNi_2O_{7-\delta}$ ($R$ = Pr, Nd, Tb) polycrystalline samples. As can be seen, the data derived from theoretical calculations and experiments can be well described by linear fitting.